\documentstyle[12pt]{article}
\begin{document}
\begin{titlepage}
\begin{flushright}
TU-606\\
gr-qc/0011043
\end{flushright}
\ \\
\ \\
\ \\
\ \\
\begin{center}
{\LARGE \bf
Diffeomorphism on Horizon\\
as an Asymptotic  Isometry \\
of Schwarzschild Black Hole \\
}
\end{center}
\ \\
\ \\
\begin{center}
\Large{
M.Hotta , K.Sasaki and T.Sasaki
 }\\
{\it 
Department of Physics, Tohoku University,\\
Sendai 980-8578, Japan
}
\end{center}
\ \\
\ \\

\begin{abstract}
It is argued that the diffeomorphism on the horizontal sphere can be regarded
 as a nontrivial asymptotic isometry of the Schwarzschild black hole. 
 We propose a new boundary condition of asymptotic metrics 
 near the horizon and show that the condition admits  
 the local time-shift and diffeomorphism on the horizon 
 as the asymptotic symmetry.
\end{abstract}

\end{titlepage}

\section{
Introduction
}
\ \\

Recently  a possibility  of 
 nontrivial  asymptotic isometry near the black hole horizon
 has attracted much attention 
in the context of the microstate counting problem \cite{HI,C, AS}.
As often discussed, the no-hair theorem apparently teaches  that 
 the black hole has only three hairs; mass, angular momentum and gauge charge.
Meanwhile the black hole carries enormous amount of entropy proportional to 
the horizon area divided by the Planck area, 
that is, the Bekenstein-Hawking entropy. 
This discrepancy, the black hole entropy problem, 
has annoyed a lot of theorists  for long time.
Among various attempts for the problem, 
 Carlip \cite{C} advocated recently  
 that by virtue of the horizon structure,  
 the ``would-be gauge freedom" of general covariance 
 supplies  physical states 
 of  the Kerr black holes in  arbitrary dimensions. 
 Those states are considered as 
 additional hairs  contributing to the entropy.
 He argued that the Kerr black holes enjoy  Virasoro horizontal isometries 
 with non-vanishing central charge and the corresponding 
  Bekenstein-Hawking entropies 
  can be reproduced by state-counting via the Cardy formula 
  in the two-dimensional conformal field theory. 
 Unfortunately there are 
  some shortcomings \cite{M} in his analysis. Thus  
 the  complete resolution of the problem has not yet been achieved. 
 However his conceptual idea 
 has  stimulated many people's interest 
 on the analysis of the horizontal isometry.

In this paper 
we propose a new  horizontal  isometry which was never discussed so far. 
It is argued that the diffeomorphism on the horizon can be regarded
 as a nontrivial asymptotic isometry of the Schwarzschild black hole. 
 Our boundary condition of the asymptotic metric is formulated in a general 
 framework.
  and show that the asymptotic condition admits the non-vanishing 
  charge of the diffeomorphism.
 This result may be significant to resolve the entropy problem.
 The analysis is performed for the black hole in four dimensions 
 in this paper, 
 but the generalization to the system in higher dimensions is straightforward.

\section{Diffeomorphism on Horizon as Asymptotic Isometry }
\ \\

In this section we propose a boundary condition  of the asymptotic metrics 
 under which  
 the diffeomorphism on the horizon can be regarded 
  as a subgroup of the asymptotic isometry.  
Let us start from the Schwarzschild metric in four dimensions.
\begin{eqnarray}
d\bar{s}^2 =-\left(1-\frac{r_*}{r} \right)dt^2 +\frac{dr^2}{1-\frac{r_*}{r}}
+r^2 \left( d\theta^2 +\sin^2 \theta d\phi^2 \right).
\end{eqnarray}
Here $r_*$ denotes the radius of the horizon. 
 Introducing the tortoise coordinate;
\begin{eqnarray}
\rho=
-(r-r_*) 
-r_* \ln \left[
\frac{r-r_*}{r_*}\right],
\end{eqnarray}
the metric near horizon  is expressed as
\begin{eqnarray}
d\bar{s}^2 &=& e^{-\frac{\rho}{r_*}}
\left[1-2e^{-\frac{\rho}{r_*}}+\cdots \right]
\left(-dt^2 +d\rho^2 \right)
\nonumber\\
&&
+r^2_* \left[1+2 e^{-\frac{\rho}{r_*}} +\cdots \right] 
\left( d\theta^2 +\sin^2 \theta d\phi^2 \right).
\label{2.1}
\end{eqnarray}
Here the horizon at $r=r_*$  is mapped into $\rho=\infty$.
Note that the leading term of the $t-\rho$ part of eqn(\ref{2.1})
  coincides with the Rindler metric. Thus, as known well, it shows
 a scaling behavior.  Consider a scale transformation
 for the metric in eqn(\ref{2.1}): 
\begin{eqnarray}
&&
t'=\frac{\epsilon}{r_*} t,
\\
&&
\rho'=\frac{\epsilon}{r_*}
\left[
\rho +2r_* \ln \left(\frac{\epsilon}{r_*}\right)
\right],\label{2.3}
\end{eqnarray}
where $\epsilon$ is a constant.
 The transformation  yields a similar asymptotic metric 
 in which $\epsilon$ appears as its length unit in the $t-\rho$ part: 
\begin{eqnarray}
d\bar{s}^2 &=& 
e^{-\frac{\rho'}{\epsilon}}
\left[
1-2
\left(\frac{\epsilon}{r_*} \right)^2 
e^{-\frac{\rho'}{\epsilon}} 
+\cdots
\right](-dt^{'2} +d\rho^{'2} )
\nonumber\\
&&
+r^2_* \left[
1 +
2\left(\frac{\epsilon}{r_*} \right)^2 
e^{-\frac{\rho'}{\epsilon}} 
+\cdots
\right]
\left(d\theta^2 +\sin^2 \theta d\phi^2 \right).
\label{2.4}
\end{eqnarray}
Now let us introduce the boundary condition of the asymptotic metrics 
which approach the background in 
eqn(\ref{2.4}) and discuss the  asymptotic isometry. 
 Our philosophy to select the boundary condition 
 is based on recent works \cite{C,AS}. 
 In the following argument we 
 adopt  ADM decomposition:
\begin{eqnarray}
g_{\alpha\beta}
=\left[
\begin{array}{cc}
-N^2 +N_k N^k & N_a \\
N_b & h_{ab}
\end{array}
\right].
\end{eqnarray}
Assume that the lapse and shift functions behave as
\begin{eqnarray}
&&
N =e^{-\frac{\rho}{2\epsilon}} 
+O\left(e^{-\frac{3\rho}{2\epsilon}} \right),
\label{2.5}
\\
&&
N^\rho =O\left(e^{-\frac{\rho}{\epsilon}} \right),
\\
&&
N^\theta =O\left(e^{-\frac{\rho}{\epsilon}} \right),
\\
&&
N^\phi =O\left(e^{-\frac{\rho}{\epsilon}} \right).
\label{2.6}
\end{eqnarray}
Also assuming that $\xi^\mu =O(1)$, it is noticed
 due to eqns(\ref{2.5})$\sim$(\ref{2.6}) that 
the leading term of the surface deformation vector is fixed 
and does not depend on 
 how the metric approaches the background as follows.
\begin{eqnarray}
&&
\hat{\xi}^t = N\xi^t \sim e^{-\frac{\rho}{2\epsilon}} \xi^t ,
\label{2.27}
\\
&&
\hat{\xi}^a =\xi^a + N^a \xi^t \sim \xi^a.
\end{eqnarray}
For the spatial components of the metric
  the following boundary conditions are adopted.
\begin{eqnarray}
&&
h_{\rho\rho} =e^{-\frac{\rho}{\epsilon}} 
+O\left(e^{-2\frac{\rho}{\epsilon}} \right),\label{2.8}
\\
&&
h_{\rho\theta} =O\left(e^{-2\frac{\rho}{\epsilon}} \right),
\label{2.9}
\\
&&
h_{\rho\phi} =O\left(e^{-2\frac{\rho}{\epsilon}} \right),
\label{2.10}\\
&&
h_{\theta\theta} =O(1),
\label{2.11}\\
&&
h_{\theta\phi} =
O(1),
\label{2.12}\\
&&
h_{\phi\phi} =O(1).
\label{2.13}
\end{eqnarray}
Later we  concentrate on the subset of those metrics 
which boundaries at $\rho=\infty$ 
 are not any physical singular surfaces. Thus  
 the curvatures of the metrics are finite on the boundary even though a part of the metric components possess the superficial coordinate singularities:
\begin{eqnarray}
&&
|R_{\mu\nu\lambda\xi} R^{\mu\nu\lambda\xi}(\rho=\infty)| <\infty ,\\
&&
|R_{\mu\nu}R^{\mu\nu}(\rho=\infty) |<\infty,
\\
&&
|R(\rho= \infty)|<\infty.
\end{eqnarray}

It is worth noting here that the leading behavior of the metric in 
eqns (\ref{2.5}) $\sim$ (\ref{2.6}) and (\ref{2.8})$\sim$(\ref{2.10})
 preserves  the horizon structure
 even for the asymptotic geometries,
  as long as the geometries are regular at $\rho =\infty$. 
Introducing the following global coordinates:
\begin{eqnarray}
&&
T = 2\epsilon e^{-\frac{\rho}{2\epsilon}}
\sinh\left( \frac{t}{2\epsilon}\right),
\\
&&
X = 2\epsilon e^{-\frac{\rho}{2\epsilon}}
\cosh\left( \frac{t}{2\epsilon}\right),
\end{eqnarray}
the metric is reexpressed as
\begin{eqnarray}
ds^2 &=& -dT^2 +dX^2 \nonumber\\
&&+O(X-T)(dX+dT)d\theta+O(X+T) (dX-dT) d\phi
\nonumber\\
&&+O((X -T)^2 ) (dX +dT )^2
\nonumber\\
&&
+O((X+T)(X-T) )(dX +dT)(dX -dT )
\nonumber\\
&&
+O((X+T)^2 ) (dX-dT)^2
\nonumber\\
&&
+O(1)d\theta^2 +O(1) d\theta d\phi +O(1) d\phi^2.
\label{555}
\end{eqnarray}
In this coordinate the boundary at $\rho= \infty$ is mapped into  $\pm T=X >0$.
Clearly the fall-off condition in eqn(\ref{555}) guarantees
  that worldlines with which the boundary is completely covered: 
\begin{eqnarray}
&&
T=X>0  \ (or -T=X >0), \label{1001}\\
&&
\theta =const ,\\
&&
\phi =const\label{1002}
\end{eqnarray}
 are exactly null geodesics 
 for every asymptotic metric. Therefore
  the horizontal structure is really exposed, 
  that is,  any physical object across the boundary ($\rho=\infty$) cannot come back to the outside region ($\rho<\infty$).

 Strictly speaking, we  call the boundary ``horizon" 
  just in the sense of the Rindler wedge. The wedge structure  
 has often been argued  
 to be very essential for the black hole thermodynamics 
 from the early stage of investigation on the Hawking radiation \cite{bd}. 
 This is mainly because  the thermal noise
 of quantum fields appears in the wedge(Rindler) coordinates.
 We believe that our boundary condition which involves the wedge structure 
 is rather natural to apply to the black hole entropy problem.

 So far we have mentioned a horizontal interpretation of our boundary 
   in a rather geometrical way 
  by introducing the boundary null geodesics.  
  If more coordinate-independent expression of 
  our fall-off metric condition near the horizon can be attained, 
  it will help us  understand more clearly
 the geometrical meaning of the boundary condition.
  We think that it is  a significant open problem.

In our formulation we  can analyze simultaneously  
 the Schwarzschild black holes with various radii larger than $\epsilon$. 
 By choosing the coordinates adequately, 
 the metrics of the black holes belong  to the same class 
 of asymptotic equivalence.    
Also stress  that 
 the constant $\epsilon$ is able to 
  be selected arbitrarily in the classical gravity.
 For each value of $\epsilon$,  
 one can always span a coordinate frame in which the leading terms of 
 the metrics of the black holes with $ r_* > \epsilon$ 
 are given by eqn(\ref{2.4}).
  If it is possible to observe the metric deviations near the horizon, 
  it may be useful for the analysis to fix the constant $\epsilon$ 
  of order of the device scale. 
On the other hand, in the microstate counting of the black hole, it might be crucial to choose 
 the Planck length for the constant $\epsilon$.

The asymptotic behaviors of 
the inverse spatial metric is obtained straightforwardly
 from eqns (\ref{2.8}) $\sim$(\ref{2.13}) as follows. 
\begin{eqnarray}
[ h^{ab} ]
=
\left[
\begin{array}{ccc}
e^{\frac{\rho}{\epsilon} }+O\left(1\right) & 
O\left(e^{-\frac{\rho}{\epsilon} }\right) &
O\left(e^{-\frac{\rho}{\epsilon} }\right) \\
O\left(e^{-\frac{\rho}{\epsilon} }\right) &
 O\left(1\right) & O\left(1\right) \\
O\left(e^{-\frac{\rho}{\epsilon} }\right) & O\left(1\right) & O\left(1\right)
\end{array}
\right].\label{2.14}
\end{eqnarray}
Note that the leading term of $h^{\rho\rho}$
always coincides with the background term; $e^{\frac{\rho}{\epsilon} }$.
The boundary conditions of the connections are also easily found using
eqns (\ref{2.8}) $\sim$(\ref{2.14}) .
\begin{eqnarray}
&&
\Gamma^{(3)\rho}_{\rho\rho} = O\left(1\right),
\\
&&
\Gamma^{(3)\alpha}_{\rho\rho} = O\left(e^{-2\frac{\rho}{\epsilon} }\right),
\\
&&
\Gamma^{(3)\rho}_{\alpha\rho} = O\left(e^{-\frac{\rho}{\epsilon} }\right),
\\
&&
\Gamma^{(3)\rho}_{\alpha\beta} = O\left(1\right),
\\
&&
\Gamma^{(3)\alpha}_{\rho\beta} = O\left(e^{-\frac{\rho}{\epsilon} }\right),
\\
&&
\Gamma^{(3)\gamma}_{\alpha\beta} = O\left(1\right),
\end{eqnarray}
where $\alpha$,$\beta$ and $\gamma$ take the angular indices  
$\theta$ and $\phi$. Under the conditions in 
eqns(\ref{2.5}) $\sim$(\ref{2.6}) and (\ref{2.8})$\sim$(\ref{2.13}), 
it is checked that
 the extrinsic curvature;
\begin{eqnarray}
K_{ab} =\frac{1}{2N}(N_{a|b}+N_{b|a} -\partial_t h_{ab}  )
\end{eqnarray}
behaves near the horizon as
\begin{eqnarray}
&&
K_{\rho\rho} =O\left(e^{-\frac{3\rho}{2\epsilon}} \right),
\label{2.15}\\
&&
K_{\rho\theta} =O\left(e^{-\frac{\rho}{2\epsilon}} \right),
\\
&&
K_{\rho\phi} =O\left(e^{-\frac{\rho}{2\epsilon}} \right),
\\
&&
K_{\theta\theta} =O\left(e^{\frac{\rho}{2\epsilon}} \right),
\\
&&
K_{\theta\phi} =O\left(e^{\frac{\rho}{2\epsilon} }\right),
\\
&&
K_{\phi\phi} =O\left(e^{\frac{\rho}{2\epsilon} }\right).
\label{2.16}
\end{eqnarray}
Here it has been assumed that 
$\partial_t h_{ab}$ shows  
the same asymptotic behavior of $h_{ab}$ near the horizon, 
except $h_{\rho\rho}$. 
For the $(\rho\rho)$ component, 
we assume from eqn(\ref{2.8}) that $\partial_t h_{\rho\rho} =O\left(e^{-2\frac{\rho}{\epsilon}} \right)$ holds.  
As usual, the canonical momentum tensor is defined as
\begin{eqnarray}
\Pi^{ab} =\frac{\sqrt{h}}{16\pi G}
[Kh^{ab} -K^{ab}],
\end{eqnarray}
and behaves under the condition near the horizon as
\begin{eqnarray}
&&
\Pi^{\rho\rho} =O\left(e^{\frac{\rho}{\epsilon}}  \right),
\label{2.28}
\\
&&
\Pi^{\rho\alpha} =O\left(1 \right),
\\
&&
\Pi^{\alpha\beta} =O\left(1 \right).
\label{2.29}
\end{eqnarray}

It is quite significant to emphasize that the following 
transformation  possesses the well-defined  charges
 and can be regarded as a subgroup of the asymptotic isometry.
\begin{eqnarray}
&&
\xi^t = T (\theta,\phi)+O\left(e^{-\frac{\rho}{\epsilon}} \right),
\label{2.25}
\\
&&
\xi^\rho = 0,
\\
&&
\xi^\theta =V^\theta (\theta ,\phi)+O\left(e^{-2\frac{\rho}{\epsilon}} \right)
,\label{2.26'}
\\
&&
\xi^\phi =V^\phi (\theta ,\phi) +O\left(e^{-2\frac{\rho}{\epsilon}} \right),
\label{2.26}
\end{eqnarray}
where $T$ and $V^\alpha$ are regular scalar and vector functions on the 
horizontal sphere. The vector part $V^\alpha$ consists of the 
 the diffeomorphism on the sphere.

Let us examine the canonical charges  explicitly.
The canonical Hamiltonian and momentum densities are written down as 
\begin{eqnarray}
&&
{\cal H}_t
=
\frac{16\pi G}{\sqrt{h}}
\left[
\Pi^{ab}\Pi_{ab}
-\frac{1}{2}\Pi^2
\right]
-\frac{\sqrt{h}}{16\pi G}
R^{(3)} +{\cal H}^{matter}_t,
\label{2.20}
\\
&&
{\cal H}_a =
-2\Pi_{ab} |^b +{\cal H}^{matter}_a .\label{2.21}
\end{eqnarray}
where $G$ is the gravitational constant, $R^{(3)}$ is the scalar curvature of 
the spatial section and ${\cal H}^{matter}_\mu$ represents 
 matter contribution of the model if it exists. 
Then the bulk part of the canonical generator is expressed 
using eqns (\ref{2.20}) and (\ref{2.21}) as
\begin{eqnarray}
H[\hat{\xi}]=\int d^{3} x \left[
\hat{\xi}^t {\cal H}_t
+
\hat{\xi}^a {\cal H}_a
\right].
\end{eqnarray}
Taking variations of $H$ with respect to $ h_{ab}$ and $\Pi^{ab}$
 yields 
\begin{eqnarray}
\delta H[\hat{\xi}]=
\int d^{3} x 
\left( 
\partial_t \Pi^{ab} \delta h_{ab}
-\partial_t h_{ab}\delta \Pi^{ab}
\right)-\delta Q,
\end{eqnarray}
where the equations of motion have been used to get the time derivative terms 
and 
\begin{eqnarray}
\delta Q
&=&\int d\phi d\theta 
\left[
\frac{1}{16\pi G}\sqrt{h}(h^{ac}h^{b\rho} -h^{ab} h^{c\rho} )
(\hat{\xi}^t \delta h_{ab|c} -\hat{\xi}^t_{|c}\delta h_{ab} )
\nonumber\right.\\
&&
\left.
\ \ \ \ \ \ \ \ \ \ \ \ \ \ 
+2\hat{\xi}^a \delta \Pi_a^\rho -\hat{\xi}^\rho \Pi^{ab}\delta h_{ab}
\right]_{\rho=\infty}.
\label{2.23}
\end{eqnarray}
If $\delta Q$ in eqn (\ref{2.23}) can be integrated rigorously,
 the canonical charge of the surface deformation is defined by
 $Q=\int \delta Q$.  Interestingly, under the conditions in 
 eqns(\ref{2.27})$\sim$(\ref{2.13}) and (\ref{2.28})$\sim$(\ref{2.29})
 it is proven that the integration of $\delta Q$ 
  for the transformation in eqns (\ref{2.25}) $\sim$ (\ref{2.26}) 
  is really possible.
The non-vanishing contribution in the limit of $\rho\rightarrow \infty$
 turns out to be just two terms as follows.
\begin{eqnarray}
\delta Q
=\int d\phi d\theta 
\left[
-
\frac{1}{32\pi G\epsilon} \sqrt{\det [h_{\alpha\beta}]}
\xi^t h^{\alpha\beta}\delta h_{\alpha\beta} 
+2\xi^\alpha \delta \Pi_\alpha^\rho 
\right]_{\rho=\infty}.
\end{eqnarray}
Here  $\alpha$ and $\beta$ run just in the angular 
 part $\theta$ and $\phi$.  Due to the relation:
\begin{eqnarray}
\delta \sqrt{\det [h_{\alpha\beta} ]}=\frac{1}{2}
\sqrt{\det[ h_{\alpha\beta}] } h^{\alpha\beta}\delta h_{\alpha\beta},
\end{eqnarray}
 the integration results in
\begin{eqnarray}
Q
=\int d\phi d\theta 
\left[
-
\frac{1}{16\pi G \epsilon} \xi^t \sqrt{\det [h_{\alpha\beta}]}
+2\xi^\alpha  \Pi^{\rho}_\alpha 
\right]_{\rho=\infty } .
\label{2.33}
\end{eqnarray}
Both two right-hand-side terms in eqn (\ref{2.33}) can be shown finite 
under the asymptotic conditions.
Also we are  able to rewrite the expression of $Q$  by use of the original 
surface deformation vector $\hat{\xi}^\mu $ as follows.
\begin{eqnarray}
Q
=\int d\phi d\theta 
\left[
\frac{\sqrt{h}}{8\pi G } h^{\rho\rho} \partial_\rho \hat{\xi}^t  
+2\hat{\xi}^\alpha \Pi^{\rho}_\alpha 
\right]_{\rho=\infty },
\end{eqnarray}
where we have used the following asymptotic forms:
\begin{eqnarray}
&&
\sqrt{h}\sim e^{-\frac{\rho}{2\epsilon}}\sqrt{\det[h_{\alpha\beta}]},
\nonumber\\
&&
h^{\rho\rho} \sim e^{\frac{\rho}{\epsilon}},
\nonumber\\
&&N\sim e^{-\frac{\rho}{2\epsilon}}.\nonumber
\end{eqnarray}
In Appendix it is commented that the canonical charge in eqn (\ref{2.33})
 coincides explicitly with the Noether charge of the transformation in
 eqns (\ref{2.25}) $\sim$ (\ref{2.26}). 

In the next section we calculate the charges of black hole solutions 
and analyze the representations.

\section{Horizontal Charge of Black Hole Solutions}
\ \\
 
In this section we evaluate the horizontal charges of the black hole solutions.
For convenience of later explanation, 
let us first introduce an angular metric on a unit sphere as follows.
\begin{eqnarray}
d\Omega^2 &=& \sigma_{\theta\theta}d\theta^2
+2\sigma_{\theta\phi}d\theta d\phi 
+\sigma_{\phi\phi}d\phi^2
\nonumber\\
&=& d\theta^2 +\sin^2 \theta d\phi^2 .
\label{3.1}
\end{eqnarray}
In general it is possible to decompose the regular vector field $V^\alpha$ 
in eqns(\ref{2.26'}) and (\ref{2.26}) into sum of a divergenceless 
(or area preserving) part and a rotationless part as
\begin{eqnarray}
V^\alpha =-\frac{1}{\sqrt{\sigma}}\epsilon^{\alpha\beta} \nabla^{(2)}_\beta
 \Psi_1 (\theta ,\phi)
+\nabla^{(2)\alpha} \Psi_2 (\theta ,\phi),
\end{eqnarray}
where $\Psi_k$ is a regular scalar function on the sphere, 
$\nabla^{(2)}$ is the covariant derivative associated with 
the metric in eqn (\ref{3.1}) and
\begin{eqnarray}
&&
\epsilon^{\theta\phi}=-\epsilon^{\phi\theta} =1,
\\
&&
\epsilon^{\theta\theta}=\epsilon^{\phi\phi} =0,
\\
&&
\sqrt{\sigma}=\sin\theta.
\end{eqnarray}
Now we have three types of  the generators of the transformations  
in eqns (\ref{2.25}) $\sim$(\ref{2.26}); $T$, $\Psi_1$ and $\Psi_2$
 and can expand them using the spherical harmonic functions as follows.
\begin{eqnarray}
&&
T=\sum_{lm} a_{lm} Y_{lm} (\theta,\phi),
\label{3.2}\\
&&
\Psi_k =\sum_{lm} b^{(k)}_{lm} Y_{lm}(\theta ,\phi).
\label{3.3}
\end{eqnarray}
Consequently all the independent generators of the symmetry 
 can be  listed as
\begin{eqnarray}
&&
J^{(t)}_{lm} = Y_{lm} \partial_t +O\left(e^{-\frac{\rho}{\epsilon}} \right),
\\
&&
J^{(1)}_{lm} = 
\frac{1}{\sin\theta}\left(
\partial_\theta Y_{lm} \partial_\phi
-
\partial_\phi Y_{lm}  \partial_\theta
\right)+O\left(e^{-2\frac{\rho}{\epsilon}} \right),
\\
&&
J^{(2)}_{lm} =\partial_\theta Y_{lm} \partial_\theta
+\frac{1}{\sin^2 \theta} \partial_\phi Y_{lm} \partial_\phi
+O\left(e^{-2\frac{\rho}{\epsilon}} \right),
\end{eqnarray}
where $J^{(t)}$ generates the local time-shift on the horizon. 
The charges $J^{(1)}$ and $J^{(2)}$ 
generate the two-dimensional diffeomorphism on the sphere. 
In order to analyze the commutation relations of the generators,
it is useful to define three structure coefficients $G_{lml'm'}^{l''m''}$,
 $C_{lml'm'}^{l''m''}$ and $D_{lml'm'}^{l''m''}$
 as
\begin{eqnarray}
&&
Y_{lm} Y_{l'm'} =\sum G_{lml'm'}^{l''m''} Y_{l''m''},
\\
&&
\frac{1}{\sin\theta}\left(
\partial_\theta Y_{lm} \partial_\phi Y_{l'm'}
-
\partial_\phi Y_{lm}  \partial_\theta Y_{l'm'}
\right)
=\sum C_{lml'm'}^{l''m''} Y_{l''m''},
\\
&&
\partial_\theta Y_{lm} \partial_\theta Y_{l'm'}
+\frac{1}{\sin^2 \theta}
\partial_\phi Y_{lm}  \partial_\phi Y_{l'm'}
=\sum D_{lml'm'}^{l''m''} Y_{l''m''}.
\end{eqnarray}
By definition the coefficients satisfy the following 
 permutation relations.
\begin{eqnarray}
&&
G_{l'm'lm}^{l''m''} =G_{lml'm'}^{l''m''}, 
\\
&&
C_{l'm'lm}^{l''m''} =-C_{lml'm'}^{l''m''}, 
\\
&&
D_{l'm'lm}^{l''m''} =D_{lml'm'}^{l''m''}.
\end{eqnarray}
Also the following relations  are easily proven from the definitions.
\begin{eqnarray}
&&
G_{lml'm'}^{00}=\frac{1}{\sqrt{4\pi}}\delta_{ll'}\delta_{mm'},
\\
&&
C_{lml'm'}^{00}=0,
\\
&&
D_{lml'm'}^{00}=\frac{l(l+1)}{\sqrt{4\pi}}\delta_{ll'}\delta_{mm'}.
\end{eqnarray}
Then all the commutators can be expressed by use of the coefficients.
\begin{eqnarray}
&&
[J^{(t)}_{lm},\ J^{(t)}_{l'm'} ]=0,
\\
&&
[J^{(1)}_{lm},\ J^{(t)}_{l'm'} ]=\sum C_{lml'm'}^{l''m''} J^{(t)}_{l''m''},
\\
&&
[J^{(2)}_{lm},\ J^{(t)}_{l'm'} ]=\sum D_{lml'm'}^{l''m''} J^{(t)}_{l''m''},
\\
&&
[J^{(1)}_{lm},\ J^{(1)}_{l'm'} ]=\sum C_{lml'm'}^{l''m''} J^{(1)}_{l''m''},
\\
&&
[J^{(2)}_{lm},\ J^{(2)}_{l'm'} ]=\sum C_{lml'm'}^{l''m''}
 J^{(1)}_{l''m''},
\end{eqnarray}

\begin{eqnarray}
&&
 [ J^{(1)}_{lm},\ J^{(2)}_{l'm'} ]
\nonumber\\
&=& \sum_{l''\neq 0} 
\left[
D_{lml'm'}^{l''m''}\frac{l'(l'+1) -l''(l''+1)}{l''(l''+1)}
-
G_{lml'm'}^{l''m''}\frac{l(l+1) l'(l'+1)}{l''(l''+1)}
\right] J^{(1)}_{l''m''}
 \nonumber\\
 &&
 +
 \sum_{l''\neq0} 
 C_{lml'm'}^{l''m''}\frac{l'(l'+1)}{l''(l''+1)}
 J^{(2)}_{l''m''}.
\end{eqnarray}
It is  noticed here that the Cartan subalgebra of the symmetry 
 consists of $J^{(t)}_{lm}$. All of generators $J^{(1)}_{lm}$ 
  do not commute with $J^{(t)}_{l'm'}$ except when $l=l'$ and $m=m'$ (or 
   $l'=0$ and $m'=0$). Also $J^{(2)}_{lm}$ do not commute with $J^{(t)}_{l'm'}$
  except when $l'=0$ and $m'=0$.

Substituting eqns (\ref{2.25})$\sim$ (\ref{2.26}) and 
the black hole solution (eqn (\ref{2.4})) into 
 eqn (\ref{2.33}) yields
\begin{eqnarray}
Q[T,V^\alpha]=-\frac{r_*^2 }{ 16\pi G \epsilon}
\int d\phi d\theta \sin\theta T(\theta ,\phi).
\end{eqnarray}
Therefore, using the decomposition in eqns(\ref{3.2}) and (\ref{3.3}),
we get each component of the charges as follows.
\begin{eqnarray}
&&
Q^{(t)}_{00}=-\frac{\sqrt{4\pi}r_*^2}{16\pi G \epsilon },
\label{3.5}\\
&&
Q^{(t)}_{lm} =0, \ \ (l\geq 1)
\label{3.6}
\end{eqnarray}
\begin{eqnarray}
&&
Q^{(1)}_{lm}=0,
\\
&&
Q^{(2)}_{lm} =0. 
\end{eqnarray}
It is also possible to verify that 
when the background is infinitesimally transformed by $T_1$ and $V^\alpha_1$,
 the charge $Q_2 =Q[T_2 ,V^\alpha_2]$ changes in the usual way and 
  any anomalous term like central extension is not needed:
\begin{eqnarray}
\delta_1 Q_2 
=
\frac{r_*^2 }{16\pi G \epsilon}
\int d\phi d\theta \sin\theta
\left[
V^\alpha_1 \partial_\alpha T_2
-
V^\alpha_2 \partial_\alpha T_1
\right].
\end{eqnarray}

It is easy to construct the irreducible representation of the symmetry 
 for the metrics of the black hole solutions with radius $r_*$.
The transformation of the background in eqn (\ref{2.4}) by
\begin{eqnarray}
&&
t' =t +P(\theta,\phi) +O\left(e^{-\frac{\rho}{\epsilon}} \right),
\\
&&\rho' =\rho,
\\
&&
\theta' =F(\theta ,\phi)+O\left(e^{-2\frac{\rho}{\epsilon}} \right),
\label{3.10}\\
&&
\phi'= G(\theta,\phi)+O\left(e^{-2\frac{\rho}{\epsilon}} \right)
\label{3.11}
\end{eqnarray}
yields all of the asymptotic metrics in the representation.
Here the regular function $P$ is an arbitrary local time-shift 
 on the horizontal sphere. The  functions $F$ and $G$ define an arbitrary  
regular coordinate transformation on the sphere. 
It turns out that any asymptotic metric in the representation
takes the following form.  
\begin{eqnarray}
ds^2 
&=&
e^{-\frac{\rho}{\epsilon}}
(-dt^2 +d\rho^2 )
\nonumber\\
&&
-2e^{-\frac{\rho}{\epsilon}}\left[
\partial_\theta P d\theta +\partial_\phi P d\phi 
\right]dt
\nonumber\\
&&
+
r^2_* \left[(\partial_\theta F)^2 +\sin^2 F (\partial_\theta G)^2 \right]
d\theta^2
\nonumber\\
&&
+2
r^2_* \left[\partial_\theta F  \partial_\phi F
+\sin^2 F \partial_\theta G \partial_\phi G\right]
d\theta d\phi
\nonumber\\
&&
+
r^2_* \left[(\partial_\phi F)^2 +\sin^2 F (\partial_\phi G)^2 \right]
d\phi^2
\nonumber\\
&&
+ ...
\label{3.7}
\end{eqnarray}
where $+\cdots$ denotes terms which do not contribute to the charges 
 of the symmetry.
For the metric in eqn (\ref{3.7}), the charges are evaluated as
\begin{eqnarray}
Q[T,V^\alpha]=-\frac{r_*^2}{16\pi G \epsilon }
\int d\phi d\theta \sin F
\left[
\partial_\theta F \partial_\phi G- \partial_\phi F \partial_\theta G
\right]
\left(T +V^\alpha \partial_\alpha P \right)
\label{3.100}
\end{eqnarray}
and the representation has been really proven to be non-singlet.

From the viewpoint of the black hole thermodynamics, 
asymptotic states 
which charges of the Cartan subalgebra  
take the same values of the background  
may be worth analyzing in detail.
This is because 
degeneracy of the eigenstates may supply
``hair degrees of freedom" which contributes to the Bekenstein-Hawking entropy,
as many people have already pointed out \cite{C,AS}.
However, 
opposed  to  the attempts with central extended Virasoro symmetries, 
it seems  fairly  difficult to perform microstate counting
  because of a lack of clear understanding of the quantum gravity.
   Thus let us just comment on the classical charges
 probably corresponding to the coherent states of the Cartan subalgebra.
 It can be noticed that the metrics in eqn(\ref{3.7}) which 
  $Q^{(t)}_{lm}$ are given
  by eqns(\ref{3.5}) and (\ref{3.6}) must satisfy
\begin{eqnarray}
\sin F
\left[
\partial_\theta F \partial_\phi G- \partial_\phi F \partial_\theta G
\right]=\sin\theta,
\label{3.12}
\end{eqnarray}
that is, $F$ and $G$ in eqns(\ref{3.10}) and (\ref{3.11}) 
must generate the area-preserving diffeomorphism on the sphere.
The diffeomorphism charges of the metrics satisfying eqn(\ref{3.12})
 are evaluated as follows.
\begin{eqnarray}
Q^{(1)}_{lm} =0,
\end{eqnarray}
\begin{eqnarray}
Q^{(2)}_{lm} =-\frac{r^2_*}{16\pi G \epsilon} l(l+1) 
\int d\theta d\phi \sin\theta
Y_{lm}(\theta ,\phi) P(\theta ,\phi).
\end{eqnarray}
Thus the divergenceless (area-preserving) part of the charges vanishes
and only the rotationless part is nontrivial. \\
\ \\

\section{Summary}
\ \\

In this paper it has been reported that 
 the general coordinate transformation on the horizon 
 in eqns (\ref{2.25})$\sim$
 (\ref{2.26}) must be treated as not a gauge freedom but 
  a subgroup of the asymptotic isometry. 
 Firstly, taking account of the near-horizon form of the 
 Schwarzschild solution 
 in eqn(\ref{2.4}), we have proposed a rather general condition of the  
 asymptotic metrics in eqns (\ref{2.5})$\sim$(\ref{2.6}) and 
 (\ref{2.8})$\sim$(\ref{2.13}). 
 Then it has been proven that  the non-vanishing 
  charge  of the symmetry really appears, as seen in eqn(\ref{3.100}). 
Therefore we conclude that 
 the ``would-be gauge freedom" in eqns 
 (\ref{2.25})$\sim$(\ref{2.26}) cannot be gauged away 
and thus might be relevant for the black hole entropy counting.\\
 
\ \\
\ \\
 
{\bf \large APPENDIX}
\ \\
\ \\

In this appendix we show explicitly that the canonical charge $Q$ in eqn 
(\ref{2.33}) is really the Noether charge for the transformation
in eqns (\ref{2.25}) $\sim$ (\ref{2.26}).
Let us first consider the covariant Lagrangian density for 
 a scalar matter field as an example and the gravity as follows.
\begin{eqnarray}
\sqrt{-g}{\cal L}_o =
\frac{\sqrt{-g}}{16\pi G} R +
\sqrt{-g}{\cal L}_{matter} (\phi ,\partial \phi ,g^{\alpha\beta}),
\end{eqnarray}
where ${\cal L}_{matter}$ denotes the contribution of the scalar field.
The equations of motion simply read 
\begin{eqnarray}
&&
\frac{\partial \sqrt{-g} {\cal L}_o}{\partial  g_{\alpha\beta}}
-
\partial_\mu 
\frac{\partial \sqrt{-g} {\cal L}_o}
{\partial \partial_\mu  g_{\alpha\beta}}
+
\partial_\mu \partial_\nu
\frac{\partial \sqrt{-g} {\cal L}_o}
{\partial \partial_\mu  \partial_\nu g_{\alpha\beta}} =0,
\label{a.1}\\
&& 
\frac{\partial \sqrt{-g} {\cal L}_o}{\partial  \phi}
-
\partial_\mu 
\frac{\partial \sqrt{-g} {\cal L}_o}
{\partial \partial_\mu  \phi} =0.
\label{a.2}
\end{eqnarray}
Next let us discuss the Lie derivative with respect to 
 a vector field $\xi^\mu$. 
The Lie derivatives of the metric and the scalar field are given as
\begin{eqnarray}
&&
\delta_\xi g_{\alpha\beta}
=\nabla_\alpha \xi_\beta +\nabla_\beta \xi_\alpha,
\\
&&
\delta_\xi \phi =\xi^\alpha \nabla_\alpha \phi.
\end{eqnarray}
Covariance of the Lagrangian density makes its Lie derivative form simple
 as 
\begin{eqnarray}
\delta_\xi (\sqrt{-g} {\cal L}_o ) =
\partial_\mu \left(\xi^\mu \sqrt{-g} {\cal L}_o \right).
\end{eqnarray}
By virtue of the Noether theorem, it is known that 
the equation of motion ensures the existence of conserved current :
\begin{eqnarray}
\partial_\mu \bar{J}^\mu =0,
\end{eqnarray}
where the current is defined as
\begin{eqnarray}
\bar{J}^\mu &=&
\left[
\frac{\partial \sqrt{-g} {\cal L}_o}{\partial \partial_\mu g_{\alpha\beta}}
-
\partial_\nu 
\frac{\partial \sqrt{-g} {\cal L}_o}
{\partial \partial_\mu \partial_\nu g_{\alpha\beta}}
\right]
\delta_\xi g_{\alpha\beta}
+
\frac{\partial \sqrt{-g} {\cal L}_o}
{\partial \partial_\mu \partial_\nu g_{\alpha\beta}}
\partial_\nu \delta_\xi g_{\alpha\beta}
-\xi^\mu \sqrt{-g}{\cal L}_o\nonumber\\
&&
+\frac{\partial \sqrt{-g} {\cal L}_o}{\partial \partial_\mu \phi}
\delta_\xi \phi . 
\end{eqnarray}
The spatial volume integration of $\bar{J}^0$ gives the Noether charge
 of general covariance of the Lagrangian density.
 
To incorporate the canonical Regge-Teitelboim action:
\begin{eqnarray}
\int d^4x \sqrt{-g}{\cal L}
&=& \int d^4 x\left[
\Pi^{ab} \partial_t h_{ab} -N{\cal H}_t-N^a {\cal H}_a +\sqrt{-g}
{\cal L}_{matter} 
\right]\nonumber\\
&&
-\int dt Q[\xi^t =1, \xi^a =0]
\end{eqnarray}
into the consideration,
we next subtract a total derivative term from the Lagrangian density
as  follows.
\begin{eqnarray}
\sqrt{-g}{\cal L} =\sqrt{-g}{\cal L}_o 
-\partial_\mu D^\mu (g_{\alpha\beta}, \partial_\gamma g_{\alpha\beta}).
\end{eqnarray}
where $D^\mu$ in the derivative term is fixed for 
the horizontal boundary as
\begin{eqnarray}
&&
D^0 =-\frac{\sqrt{h}}{8\pi G} K ,
\\
&&
D^\rho = \frac{\sqrt{h}}{8\pi G}
\left[
K^{\rho b}N_b -N^{|\rho}
\right]-\frac{1}{16\pi G \epsilon}\sqrt{\det [h_{\alpha\beta}]} ,
\\
&&
D^\alpha = \frac{\sqrt{h}}{8\pi G}
\left[
K^{\alpha b} N_b -N^{|\alpha}
\right].
\end{eqnarray}
In this case the Lie derivative of the Lagrangian density
 is modified by the total divergent term as
\begin{eqnarray}
\delta_\xi \sqrt{-g}{\cal L} = \partial_\mu \Lambda^\mu
\label{a.5}
\end{eqnarray}
where
\begin{eqnarray}
\Lambda^\mu =\xi^\mu \sqrt{-g}{\cal L}
+\xi^\mu \partial_\nu D^\nu -\delta_\xi D^\mu .
\end{eqnarray}
 and eqn (\ref{a.5}) yields a new current conservation law
 as follows.
\begin{eqnarray}
\partial_\mu J^\mu =0,
\end{eqnarray} 
where the current is re-defined as
\begin{eqnarray}
J^\mu &=&
\left[
\frac{\partial \sqrt{-g} {\cal L}}{\partial \partial_\mu g_{\alpha\beta}}
-
\partial_\nu 
\frac{\partial \sqrt{-g} {\cal L}}
{\partial \partial_\mu \partial_\nu g_{\alpha\beta}}
\right]
\delta_\xi g_{\alpha\beta}
+
\frac{\partial \sqrt{-g} {\cal L}}
{\partial \partial_\mu \partial_\nu g_{\alpha\beta}}
\partial_\nu \delta_\xi g_{\alpha\beta}
-\Lambda^\mu  \nonumber\\
&&
+\frac{\partial \sqrt{-g} {\cal L}}{\partial \partial_\mu \phi}
\delta_\xi \phi .
\end{eqnarray}
Let us define coefficients $B^\mu_\nu$,$C^{\mu\alpha}_\nu$ and
 $F^{\mu\alpha\beta}_\nu$ as
\begin{eqnarray}
J^\mu = B^\mu_\nu \xi^\nu 
+C^{\mu\alpha}_\nu \partial_\alpha \xi^\nu
+F^{\mu\alpha\beta}_\nu \partial_\alpha \partial_\beta \xi^\nu .
\end{eqnarray}
Then the Noether theorem tells that the following four relations
 must hold simultaneously.
\begin{eqnarray}
&&
\partial_\mu B^\mu_\nu =0,
\\
&&
B^\mu_\nu +\partial_\alpha C^{\alpha \mu}_\nu =0,
\\
&&
\frac{1}{2}\left(C^{\alpha\beta}_\nu +C^{\beta\alpha}_\nu \right)
+\partial_\mu F^{\mu\alpha\beta}_\nu =0,
\\
&&
F^{\alpha\beta\gamma}_\nu+
F^{\beta\gamma\alpha}_\nu
+F^{\gamma\alpha\beta}_\nu =0.
\end{eqnarray}
Using the above relations, the charge density $J^0$ can be reexpressed
 by a spatial total derivative term as follows.
\begin{eqnarray}
J^0 =\partial_a \left[
\partial_0 \left(F^{a00}_\mu \xi^\mu \right)
-C^{a0}_\mu \xi^\mu
-2F^{a0\nu}_\mu \partial_\nu \xi^\mu
\right].
\end{eqnarray}
Substituting the asymptotic conditions
 (\ref{2.27})$\sim$(\ref{2.13}) and (\ref{2.28})$\sim$(\ref{2.29}),
it can be shown that the Noether charge really
 equals to the canonical charge $Q$ as follows.
\begin{eqnarray}
Q_{Noether}
&=&\int d\rho d\theta d\phi J^0 
\nonumber\\
&=&-\int d\theta d\phi
\left[
C^{\rho 0}_0 \xi^t 
+
C^{\rho 0}_\alpha \xi^\alpha 
\right]_{\rho=\infty}
\nonumber\\
&=&
\int d\theta d\phi
\left[
-\frac{1}{16\pi G \epsilon }\xi^t \sqrt{\det[h_{\alpha\beta}]}
+2\xi^\alpha \Pi^\rho_\alpha
\right]_{\rho =\infty} =Q [\xi].
\end{eqnarray}

\end{document}